\documentclass[fleqn]{article}
\usepackage{espcrc2}
\usepackage{graphicx}

\newcommand{\bel}{\begin{equation}}
\newcommand{\ee}{\end{equation}}
\def\rys#1#2#3{\begin{figure}[h]
      \vskip 3mm
      \centerline
      {
      \includegraphics*[width=0.5 \textwidth]{#1}
      }
      \caption{#2}
      \label{#3}
      \vskip 3mm
      \end{figure}
      }
\def\MeV{{\rm MeV}}

\newcommand{\AmS}{{\protect\the\textfont2
  A\kern-.1667em\lower.5ex\hbox{M}\kern-.125emS}}
\title{Spectrum of recoil nucleons in quasi-elastic neutrino-nucleus interactions}
\author{Cezary Juszczak ,
Jaros\l aw A. Nowak, Jan~T.~Sobczyk\thanks{J.T. Sobczyk was
supported by KBN grant 105/E-344/SPB/ICARUS/P-03/DZ211/2003-2005;
C. Juszczak and
J.A. Nowak were supported by LNGS-TARI P10/02 }\address{Institute of Theoretical Physics, Wroc\l aw University.\\
pl. M. Borna 9, 50-204 Wroc\l aw, Poland} }

\begin{document}

\begin{abstract}
We have analyzed the consequences of introducing the local density
approximation combined with an effective nuclear
momentum-dependent potential into the CC quasi-elastic
neutrino-nucleus scattering. We note that the distribution of the
recoil nucleons momenta becomes smooth for low momentum values
and the sharp threshold is removed. Our results may be relevant
for Sci-Fi detector analysis of K2K experiments. The total amount
of observed recoil protons is reduced because some of them remain
bound inside the nucleus. We compare theoretical predictions for
a probability of such events with the results given by NUX+FLUKA
MC simulations.
\end{abstract}

\maketitle

\section{INTRODUCTION}

In recent years there has been growing interest in the studies of
neutrino interactions at energies of a few GeV \cite{NuInt}. It
was motivated by the need for more precise measurements of
neutrino oscillation parameters ($\theta_{13}$ in particular).
This entails deriving the best description of interactions with
free nucleons, and then incorporating nuclear effects. From the
point of view of the Monte Carlo codes most (or all) nuclear
effects are described with numerical packages \cite{FLUKA} but it
is enlightening how many of these effects can be presented in an
analytical form.

We investigated the process $\nu_{\mu}~n\rightarrow \mu^-~p$ with
the target neutron bound inside the nucleus. Computations of
nuclear effects were based on the Fermi gas model which is known
to work well in the above mentioned energy region \cite{FG}.  But
this approach is not completely satisfactory. For example its
simple form leads to the conclusion that the ejected nucleons can
only have momenta higher than the chosen value of the Fermi
momentum $k_F$ \cite{cut}. However there seems to be no physical
reason why lower values of momenta should be forbidden.

A possible solution to this problem is to introduce a local
density approximation \cite{LDA}. In fact the density of nuclear
matter is not uniform and accordingly it is possible to introduce
the concept of local Fermi momentum $k_F(r)$.  Since the interaction
can take place in a region where Fermi momentum is arbitrarily
low, the distribution of momenta of recoil nucleons  becomes
smoother.

Another solution can be based on a different effect.
The target nucleons are not free but they are bound
inside the nucleus. The binding energy is smaller then the typical
values of energy transfer but it may cause interesting effects.
Momentum dependent optical potential can be obtained even from
the simplest versions of nuclear mean field theory
\cite{SerotWalecka}. In the covariant approach based on the Dirac
equation one has to distinguish contributions from scalar and
vector components of nucleon self-energy in nuclear matter. In
this framework the electron-nuclei scattering was discussed in
\cite{KHF} with mean fields taken from \cite{Cooper}. More recent
results on the self-energy can be found e.g.\ in \cite{SM} where
derivations used the G-matrix approach. Our investigation is
based on other computations \cite{potential}. This choice was
dictated by two reasons. Firstly we were able to reproduce an
explicit formula for the potential as a function of density
(local Fermi momentum) and nucleon momentum. Secondly we wanted
to enable comparisons with the numerical results of other authors
who used the same potential \cite{Brieva}.

We wanted to describe both local density and potential effects and
for that we needed an analytical  form of the potential. So we
have derived a simple analytical form of the potential dependent
on two variables:  the Fermi momentum and the nucleon momentum. We
incorporated this potential into a Monte Carlo generator of events
and obtained a spectrum of the ejected nucleon momentum. The
simulations were performed for three target nuclei: oxygen, iron
and argon; these are possible targets in the neutrino experiments.
The results in all three cases were very similar because the shapes
and the density profiles did not differ significantly.

We investigated also the possibility that the effective potential leads
to reduction of the number of the produced protons that escape from the
nucleus. We compute the fraction of the total cross section which
can be interpreted as corresponding to an excited nucleus in the
final state. The same effect is predicted by other MC generators
e.g. by NUX+FLUKA.

We have also looked  for the effects related to the proton-neutron
asymmetry inside the nucleus. It turned out that the introduction
of separate values of the Fermi momentum for proton and neutron
gases left all the plots virtually unchanged.

Our results can be useful in near detector analysis of neutrino
interaction events in K2K where Sci-Fi detector registers tracks
of ejected protons. The reconstruction procedure requires the
particle to cross at least 3 planes of scintillating fiber thus
the protons would need a momentum greater than $500~\MeV$.
Therefore in the data analysis it is important to have MC which
correctly describes the shape of the distribution of recoil
protons momenta.

We performed MC simulations assuming the neutrino energy profile
expected at Sci-Fi detector \cite{spectrum}. We investigated to
what extent the spectrum of the recoil proton momentum is
influenced by the introduction of the LDA and the effective
potential.

We restricted ourselves to a simplified dynamic model without the
effects caused by the RPA correlations. We also excluded
form our analysis the effects due to the final state interactions
in nuclei.

The spectrum of the ejected nucleon momentum was studied earlier
in the context of NC reactions where it is the only observable
quantity. The outgoing nucleons with higher momentum (above the
Cherenkov threshold for SK
i.e.\ bigger than  about $1.07$~GeV) were studied
recently in \cite{beacom}. In an earlier study \cite{hor} an average
binding energy was used to describe kinematics of the process and
interesting predictions for angular distributions of recoil
nucleons were proposed. Pauli blocking was not imposed and
arbitrarily low values of the momenta of ejected nucleons were
obtained. Another approach \cite{alb} used Pauli blocking and
forbode recoil nucleons of kinetic energy lower than about
$27~\MeV$. It seems clear that any realistic model of interaction
should incorporate Pauli blocking and consequently also a
mechanism to remove the nonphysical threshold in the ejected
nucleon momentum spectrum.

\section{MODEL}

Our Monte Carlo generator simulates quasi-elastic neutrino
interactions with basic dynamics introduced according to
\cite{LS}.

The events are obtained in the following manner:

\begin{itemize}
  \item The neutrino energy $E_{\nu}$ is chosen as either a fixed value or
  generated according to some beam energy profile.

  \item The Fermi momentum is established using global or local Fermi momentum
  scheme:
  \begin{itemize}
    \item In the global scheme the Fermi momentum is fixed.

    \item In the local scheme the region in the nucleon
    where the interaction is going to take place is selected first. Then  the Fermi momentum is
    calculated based on the nuclear density in this region.
  \end{itemize}
  The actual target momentum is chosen at random from the Fermi ball of that
  radius.

  \item The nucleon neutrino pair is boosted to its center of mass frame (CMS) where the
  direction of the scattering is taken to be random but it will be weighted
  later on.

  \item While keeping this direction fixed in the CMS frame various values
  of the outgoing nucleon momentum are tested and by means of the bisection
  algorithm the value which brings about the energy conservation is chosen.
  The energy is always evaluated in the LAB frame and takes in the account
  the momentum dependent potential of the nucleon.

  \item The outgoing nucleon is Pauli blocked if its momentum in the lab is
  smaller than the local fermi momentum.

  \item The exit of the nucleon from the nuclear matter is simulated by diminishing
  the values of its momentum
  from the above calculated value $p_N$ (inside nucleus) to the value $p'$ (outside nucleus)
  calculated from the energy conservation condition
  \begin{equation}
  V(p_N)+E_k(p_N) = E_k(p')
  \end{equation}
  If there is no solution then the proton is unable to leave the
  nucleus and the nucleus becomes excited.
  The justification for this equation comes from the fact that there is no nuclear
  potential outside the nucleus.
  \item The value of cross section is calculated according to neutrino energy in the target rest frame.
  A correction of little significance (less than 1  \%
  effect on the cross section) due to nonlinearity of the neutrino and target nucleon momenta is taken into account.
  The weight of the event
  is proportional to the nuclear density, differential cross section (with correction), the boost
  jacobian, the bisection algorithm jacobian.

\end{itemize}

\rys{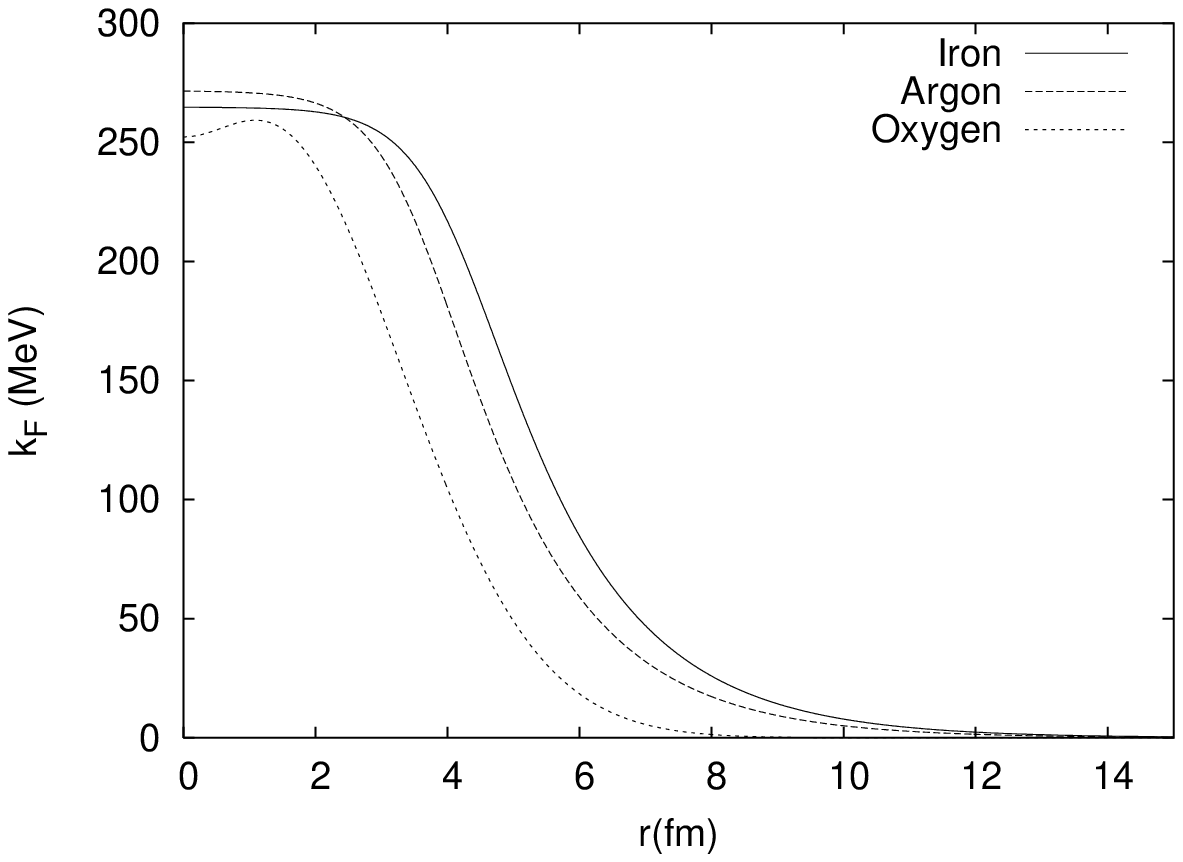}{The local Fermi momentum $r$ dependence for $_8 O^{16}$,
$_{18}Ar^{40}$ and $_{26}Fe^{56}$. }{kF}

\rys{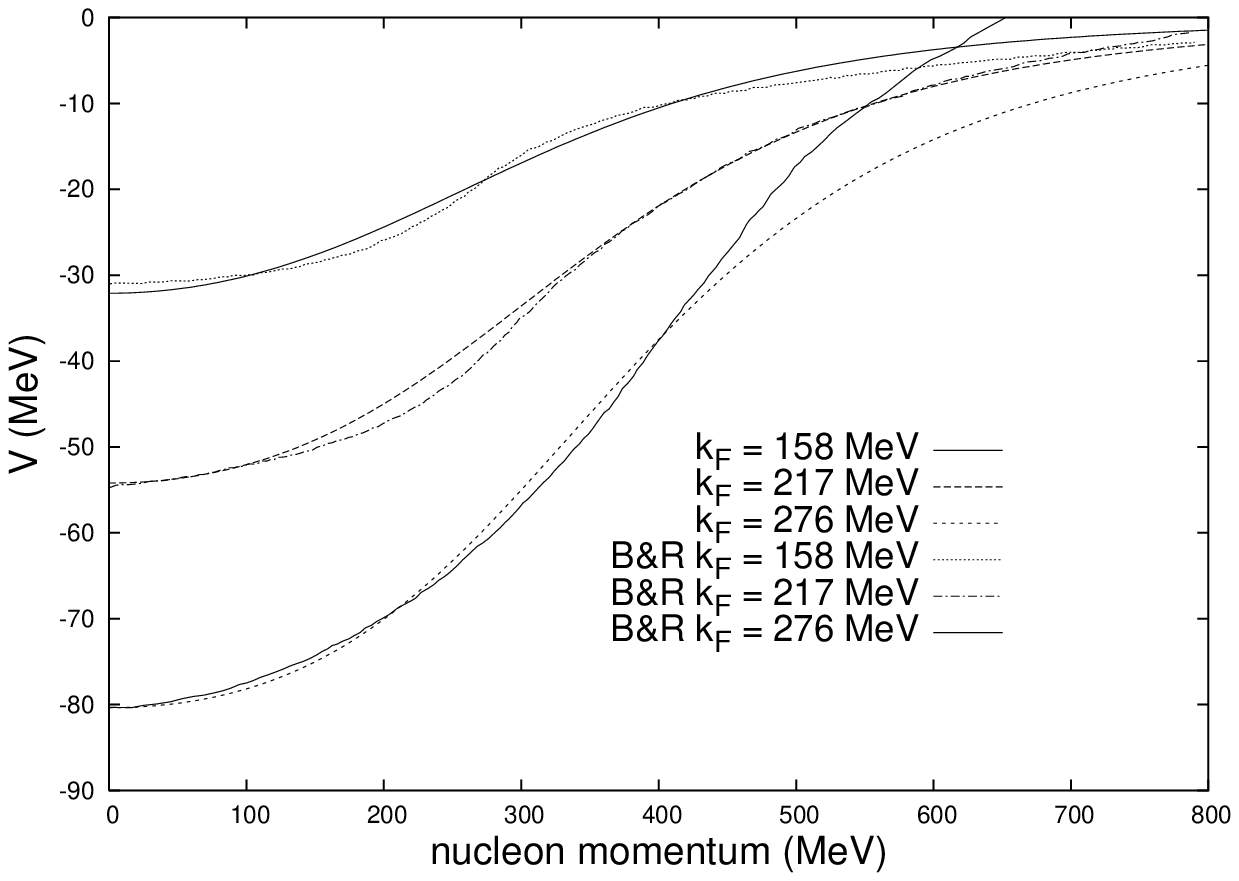}{Momentum dependent potential $ V\left( {k_F ,p}
\right)$ for 3 values of Fermi momentum (see formula (7)) compared
with three original plots taken from \cite{Brieva}.}{Vp}

    The target is treated as a collection of nucleons distributed in
space according to the density profile determined by
electron-nucleus scattering experiments. For $_8 O ^{16}$ we
adopted the harmonic oscillator model in which the nuclear density is
given by \cite{table}:

\begin{equation}
\rho^{^{16}O} \left( r \right) = \rho_0 \exp \left( { - r^2 /R^2 }
\right)\left( 1 + C{\textstyle{{r^2 } \over {R^2 }}} \right)
\end{equation}
where $R=1.883$ fm, $\rho_0 = 0.141$ fm$^{-3}$,  $C=1.544$,
and $\rho_0$ is a normalization constant defined by the condition:

\begin{equation}
\int d^3r \rho^{^{16}O} \left( r \right) = A\,.\end{equation}

For $_{18}Ar^{40}$ and $_{26}Fe^{56}$ we use the two parameter
Fermi model and write the density as:
\begin{equation}
\rho^{Ar,Fe} \left( r \right) = \frac{{\rho_ 0}}{{1 + \exp \left(
{{\textstyle{{r - C} \over {C_1 }}}} \right)}}
\end{equation}
with the following parameters:\\

\begin{tabular}{|c|c|c|c|}\hline
   & $\rho_0 [fm^{-3}]$ & $C [fm]$ & $C_1 [fm]$\\ \hline
  $_{18}Ar^{40}$ & $0.176$ & $3.530$ & $0.541$\\
  $_{26}Fe^{56}$ & $0.163$ & $4.111$ & $0.558$\\ \hline
\end{tabular}

\bigskip
The local Fermi momentum is determined by the
density profile according to:

\begin{equation}
k_F \left( r \right) = \sqrt[3]{{\frac{{3\pi ^2 \rho \left( r
\right)}} {2}}}
\end{equation}

In fig.~\ref{kF} we show the local Fermi momentum dependence on $r$ for Oxygen, Argon and Iron.
In the case of nonsymmetric nuclei one can also introduce separate
local Fermi momenta for protons and neutrons:
\begin{equation}
k_F^p(r)  = \sqrt[3]{{\frac{{2Z}} {A}}}k_F(r) , \ \ \ \ \ \  k_F^n(r) =
\sqrt[3]{{\frac{{2 {N} }} {A}}}k_F(r).
\end{equation}
where A, Z, and N are the atomic number, the number of protons, and
the number of neutrons in the nucleus respectively.

The average value of $k_F$ is calculated as:
\begin{equation}
\left\langle {k_F^{nucleus} } \right\rangle  = \frac{{\int {k_F
\left( r \right)r^2 \rho ^{nucleus} \left( r \right)dr} }} {{\int
{r^2 \rho ^{nucleus} \left( r \right)dr} }}.
\end{equation}
We get the following values:
\begin{itemize}
    \item[] $<k_F^O>=199$ \MeV,
    \item[] $<k_F^{Ar}>=217$ \MeV,
    \item[]  $<k_F^{Fe}>=217$ \MeV.\
\end{itemize}

Using tresults from \cite{potential} we find an analytic form of
the real part of the optical potential as (see fig.~\ref{Vp}):
\begin{equation}\label{potential}
 V\left( {k_F ,p} \right) = -\frac{(ak_F)^2\left(k_F+b \right)} {{c^4 +d^3k_f+e^3p^2/k_f + {p }^4 }},
\end{equation}
where $k_F$, $p$, and $V(k_F,p)$ are given in $\MeV$.
The fitted values of the parameters are:\\
 $a=206~\MeV$, $b=582~\MeV$, $c=-322~\MeV$, $d=422~\MeV$, and $e=
 289~\MeV$.

The formula (\ref{potential}) fulfills the following criteria:
\begin{itemize}
    \item V is negative
    \item V reproduces essential features of plots from \cite{potential},
    \item V is monotonously increasing,
    \item For higher values of momentum, V quickly approaches
    zero,
    \item For small values of momentum, V is proportional to
    $p^2$ as in the limit as $p\rightarrow 0$ on general ground one expects that
\begin{equation}
V(p)\sim V_0+ \frac{p^2}{2M^*}
\end{equation}
\end{itemize}

There is some amount of uncertainty in the reconstructed potential
which is inherited from the original computations in which
contributions from only finite number of harmonics were included.

\rys{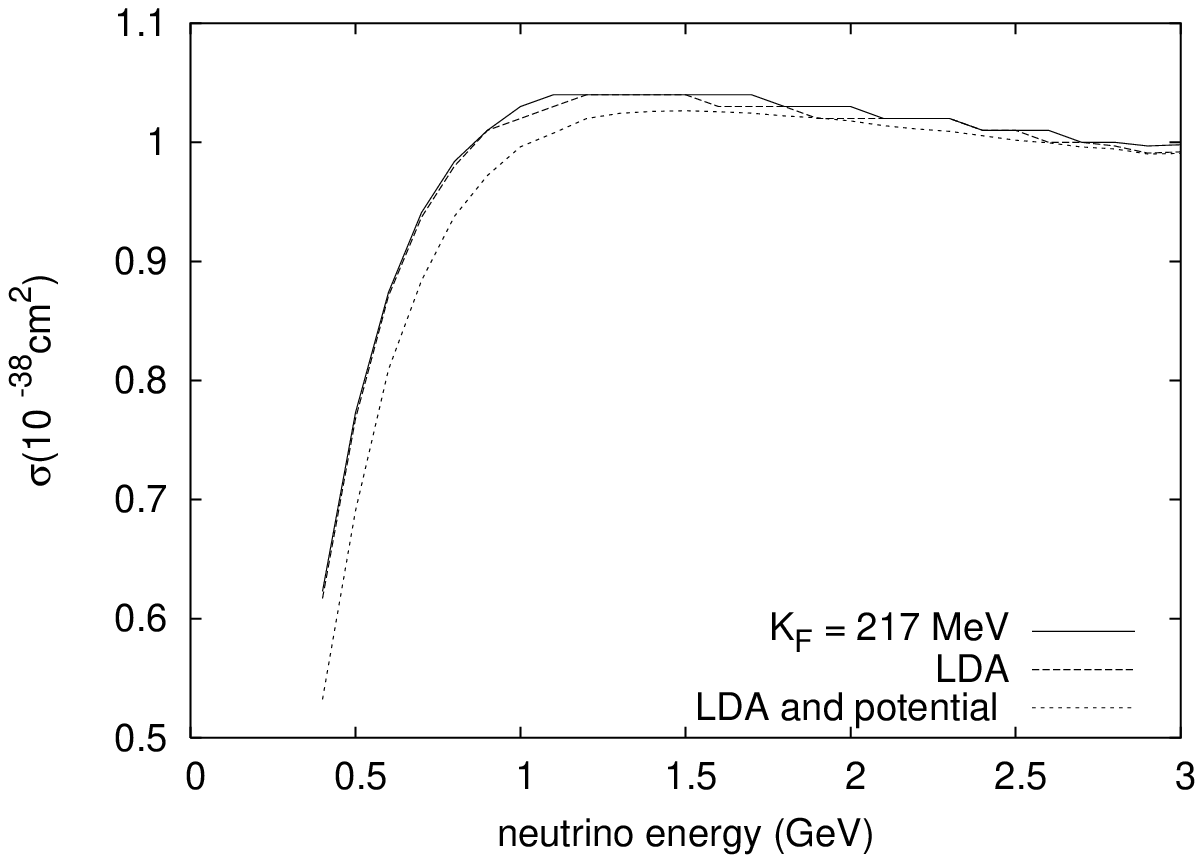}{Total cross section for quasi-elastic scattering on iron nucleus per nucleon. }{total}

\rys{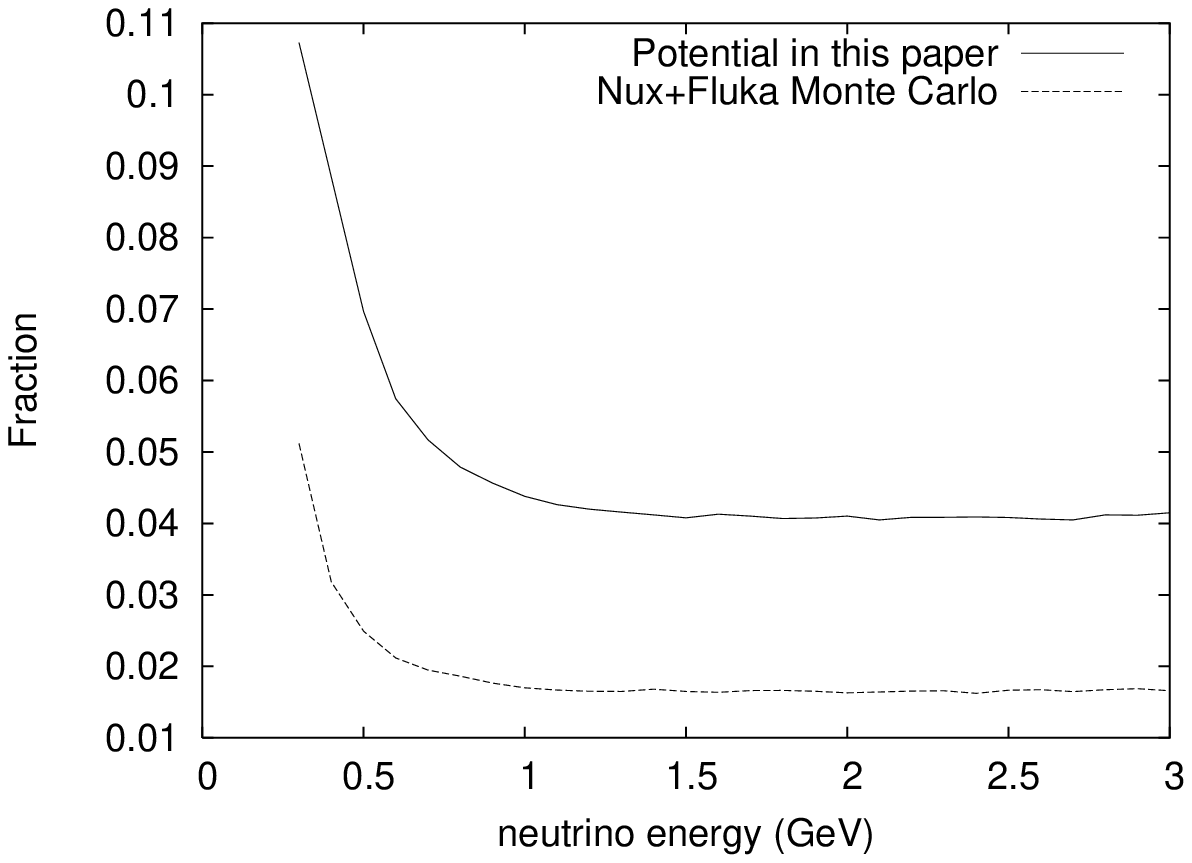}{A fraction of the total cross section
corresponding to events in which proton remains bounded in excited
nucleus.}{effect}

 \rys{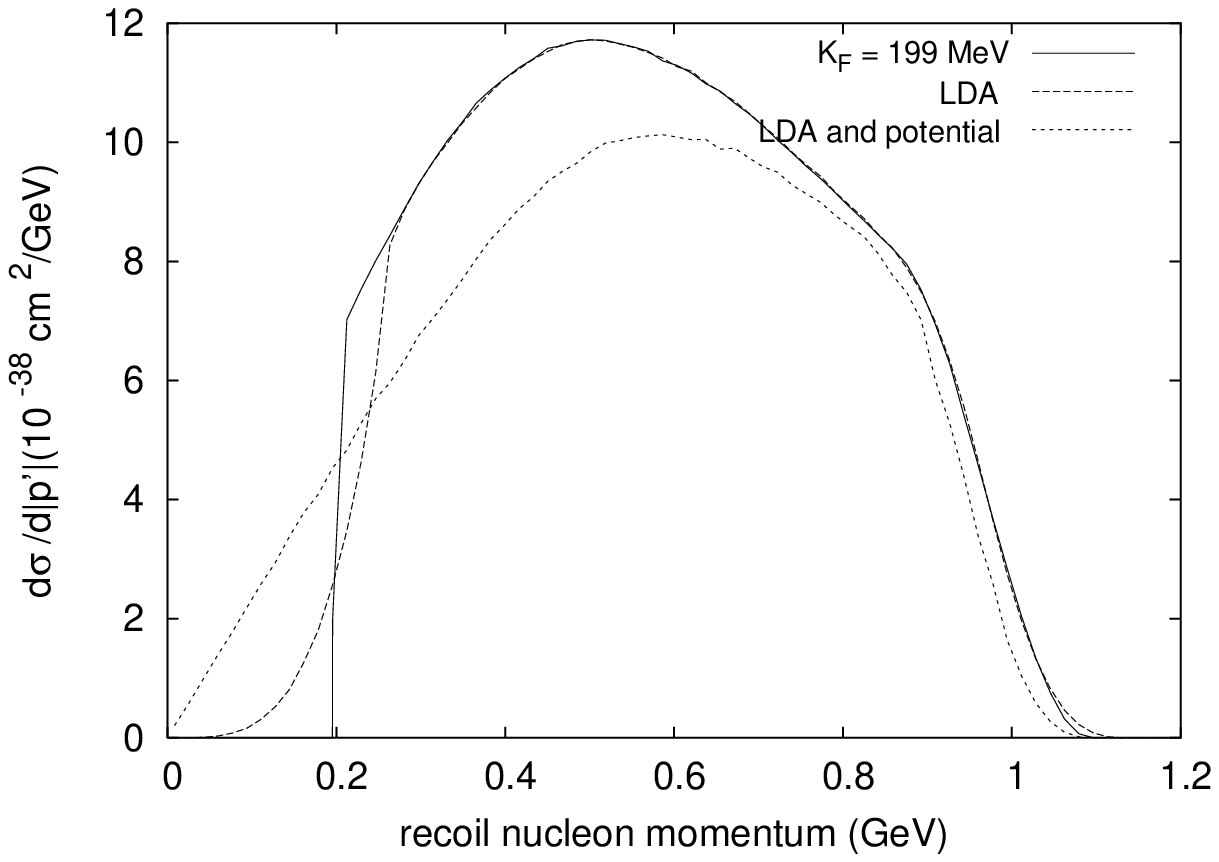}{Quasielastic neutrino-oxygen
scattering: recoil nucleons momentum distribution at $E_{\nu} =
700$ $\MeV$. }{oxygen}

\rys{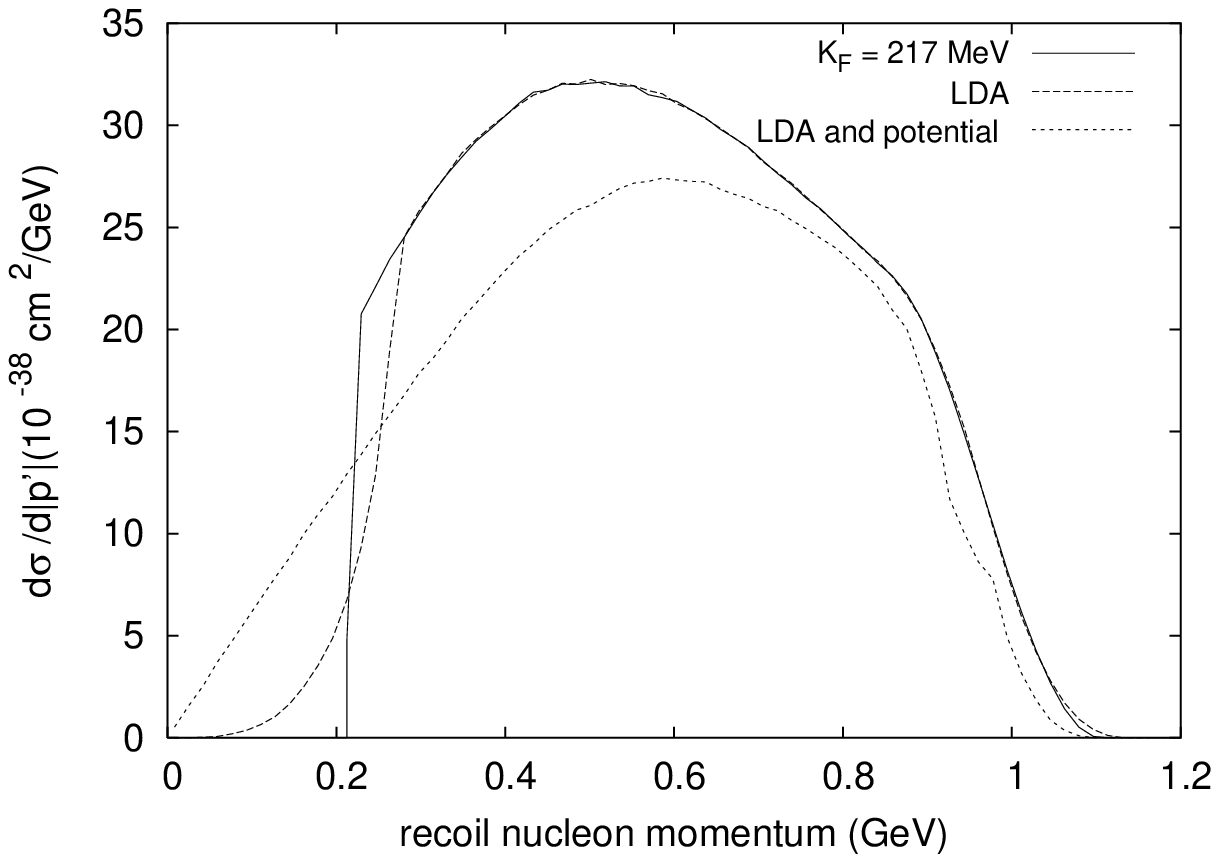}{Quasielastic neutrino-argon
scattering: recoil nucleons momentum distribution at $E_{\nu} =
700$ $\MeV$.}{argon}
 In numerical computations we compare three cases:
\begin{description}
    \item[(a)] Fermi gas with global $k_F=\left\langle {k_F^{nucleus} }
    \right\rangle$,
    \item[(b)] Fermi gas in local density approximation (LDA),
    \item[(c)] Fermi gas with in the local density approximation with momentum dependent nucleon
    potential
 (\ref{potential}).
\end{description}

\section{RESULTS}

The total cross sections for quasi-elastic interaction for all
three cases are plotted in fig.~\ref{total}. While changes
introduced by the LDA are minor, the effective potential reduces
the total cross section by a few percent. Since the produced
proton is affected by momentum depended potential it may not have
enough energy to leave the nucleus. In such a case in the final
state the nucleus is excited and there is no ejected nucleon. This
happens always if
\begin{equation}
V(p_N)+E_k(p_N) \leq M.
\end{equation}
The probability of such a process is shown in fig.~\ref{effect}. For
comparison  we calculated the probability of analogous events in
the NUX+FLUKA MC generator. It is very interesting that the
shapes of the curves are in both cases almost identical. The
probability for a bound nucleon in the final state decreases with
neutrino energy in both generators and at energy of about 1~GeV
becomes flat. However, the effect predicted by our Monte Carlo
generator is much bigger (about three times).

The predicted distributions of the ejected nucleon momentum are
shown in fig.~\ref{oxygen} and \ref{argon}. The plots are
normalized as differential cross sections ${d\sigma}/{dp'}$. In
all cases the neutrino energy is fixed at $E_\nu = 700$ $\MeV$.
The changes observed due to the LDA and the effective potential are very
similar for all three analyzed targets so we do not include a
plot for iron.

In the case (a) there are obvious sharp thresholds at $\left\langle
{k_F^{nucleus} } \right\rangle$.

In the case (b) the distributions become smoother. They differ only
in the region of nucleon momentum below $\sim 300~\MeV$; whereas,
for higher values of nuclear momentum, LDA introduces no
significant differences.

Finally, in the case (c) there is higher probability for the
outgoing nucleon to have momentum lower than $k_F$. The values
close to zero become accessible as well. The differential cross
section rises almost linearly for nucleon momenta smaller than
$450~\MeV$ then it slowly  bends reaching maximum at about
600~MeV. For momenta above 800~$\MeV$ the plot is only slightly
changed compared to  the cases (a) and (b).

The fig.~\ref{iron_kf} illustrates the relevance of the
introduction of separate values of Fermi momentum for neutron and
proton Fermi gases. The Fermi momentum for neutrons is larger and
since the interaction takes place on neutrons, it is clear that the
cross section should be slightly larger. It is indeed the case
but the effect is very small and the difference of the plots is
hardly noticeable.

\rys{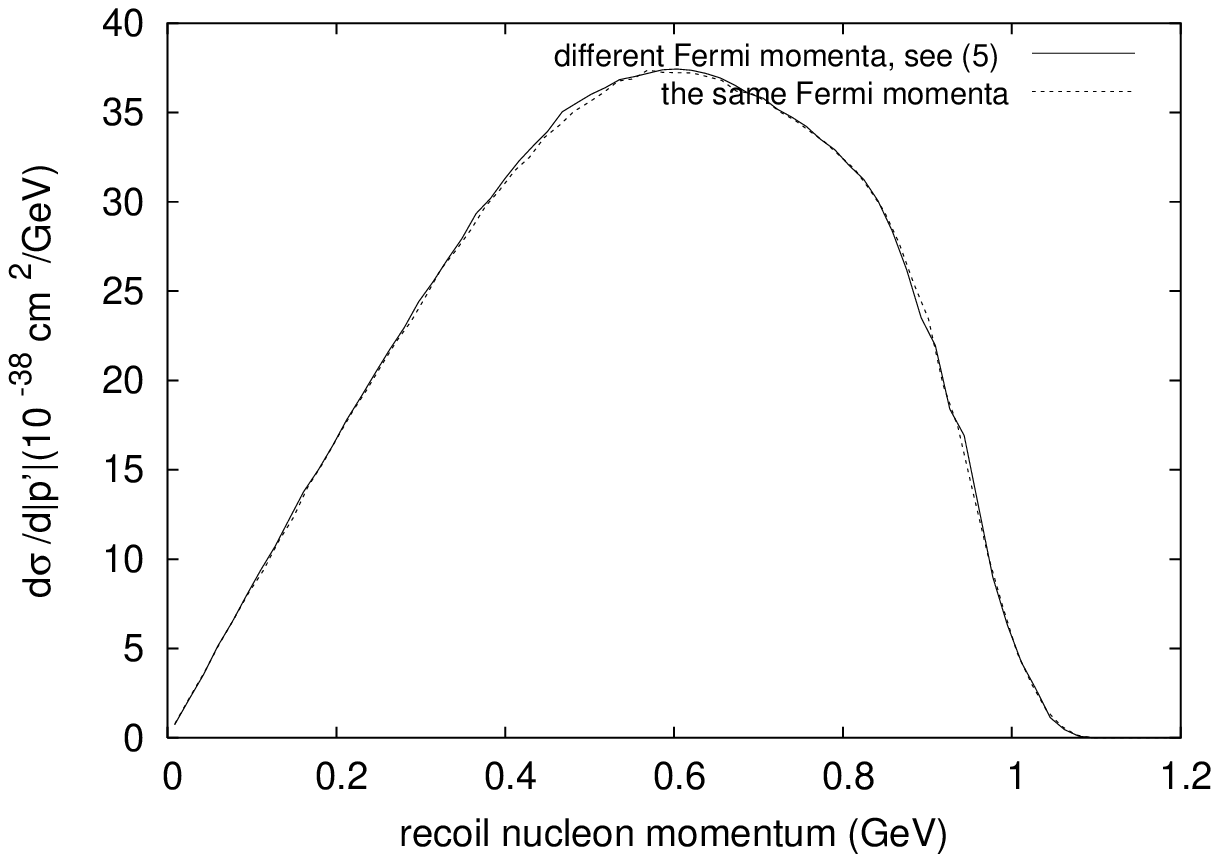}{Effect of different Fermi momenta
for proton and neutron Fermi gases in quasielastic neutrino-iron
scattering for ejected momentum distribution at $E_{\nu} =
700\MeV$.}{iron_kf}

\rys{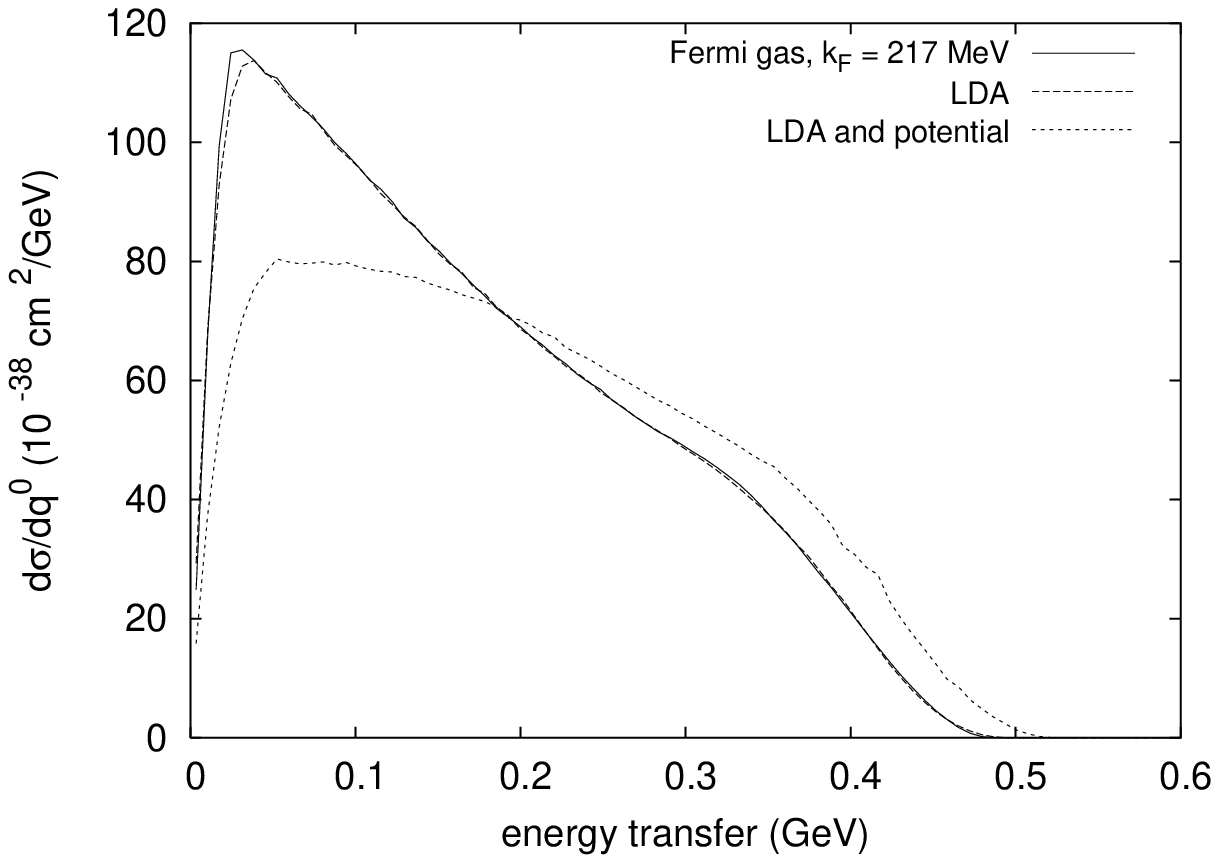}{Differential cross section of energy
transfer for quasielastic neutrino scattering on Iron; $E_{\nu} =
700\MeV$.}{iron_diff}

\rys{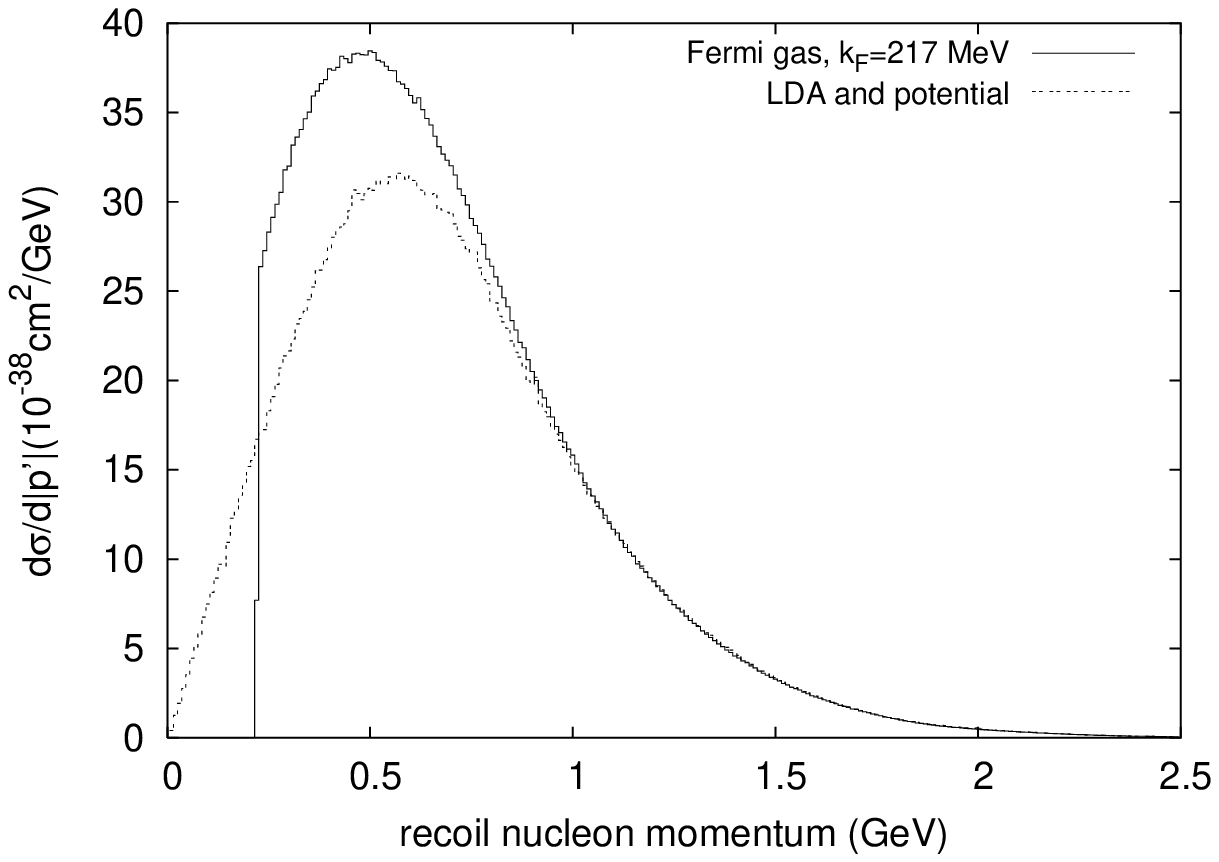}{Distribution of recoil nucleons from quasi-elastic
reactions on iron for neutrino energy spectrum identical with that
of K2K at near detector. The plot is normalized to be an effective
(energy averaged) differential cross section of ejected nucleon
momentum. LDA and potential effects reduce the total cross
section and there is also a probability of about 4\% that the
proton does not leave nucleus. In effect the area below the
corresponding curve is reduced.}{k2k_p1}

\rys{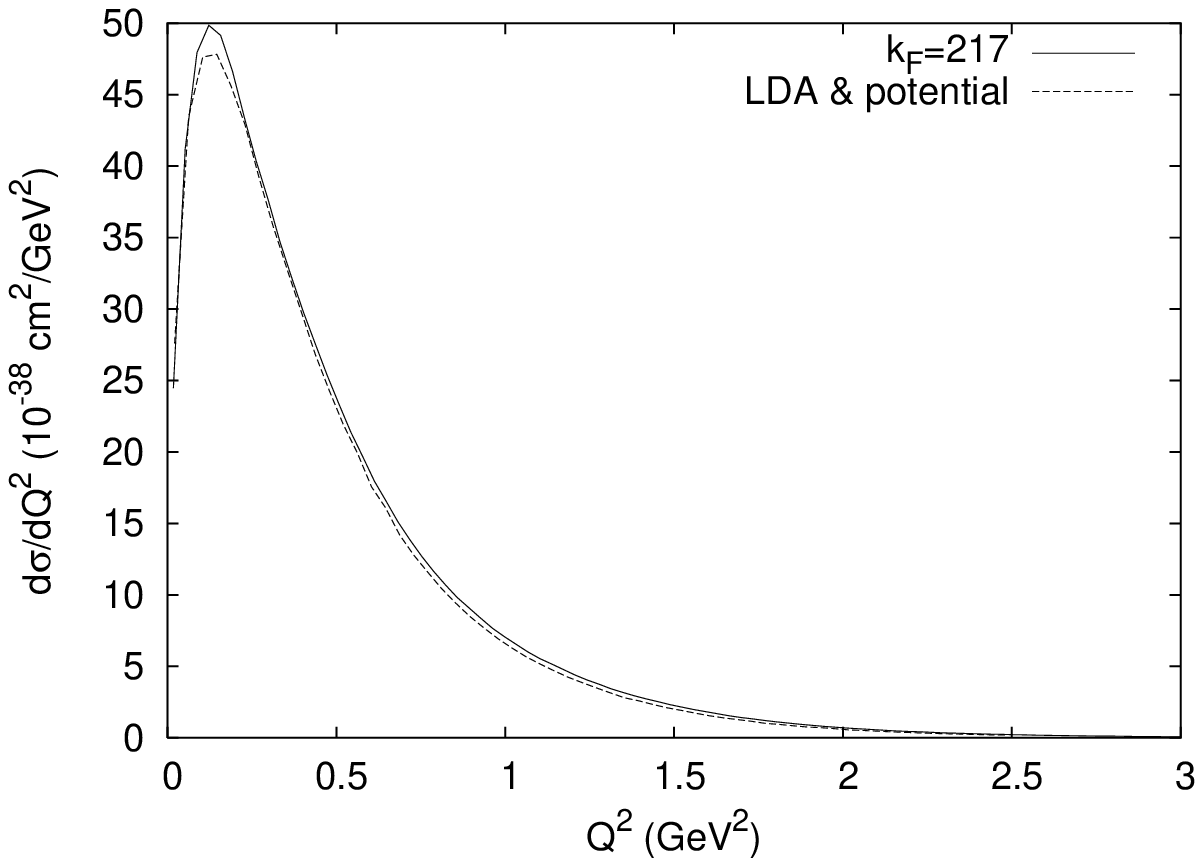}{Differential cross section of $Q^2$ transfer for
quasi-elastic reaction on iron with  K2K neutrino beam. }{k2k_q2}

For completeness in fig.~\ref{iron_diff} we present how both
discussed effects influence the energy transfer spectrum for the
iron nucleus (plots for oxygen and argon are very similar). There
is only a small difference at the peak  between the differential
cross section plots for the cases (a) and (b) the plot for the LDA
being slightly lower. In the case (c) we observe a significant
change in the shape of the plot. For lower values of the energy
transfer the differential cross section is substantially reduced
and for higher energy transfers it is slightly increased. The
allowed kinematical region is also modified.

In fig.~\ref{k2k_p1} we show the ejected proton momentum
distribution obtained for neutrino beam with the energy profile
identical to that predicted for the K2K near detector. We note
that the LDA and the potential effects remove the sharp threshold
and give a smooth plot at low values of the momentum. The number
of nucleons with momentum between  $k_F$   and  $700~\MeV$ is
significantly reduced but the values between 0 and $k_F$ become
possible.

In fig. \ref{k2k_q2} we show the energy transfer distribution for
quasi-elastic events produced with the same K2K-like beam. It is
clear that it is insensitive to the effects discussed in the present
paper.


\end{document}